\documentclass[11pt,a4paper]{article}
\usepackage{amsmath, amssymb, geometry, hyperref, booktabs}
\usepackage{graphicx} 
\usepackage{mathptmx} 
\title{Bubbles vs. Baselines: Token Valuation and Institutional Capital in PoS Networks under EIP-1559}

\author{Mikhail Perepelitsa\footnote{
Department of Mathematics, University of Houston,
\emph{PGH, 3551 Cullen Blvd, 
Houston, TX 77204-3008, USA},
\texttt{maperepelitsa@uh.edu} }
}
\date{}
\begin{document}
\maketitle

\begin{abstract}
This paper presents an open-economy macroeconomic equilibrium model for Proof-of-Stake (PoS) networks with fee-burn mechanics (EIP-1559) that formalizes the strategic interplay between a Kelly-optimizing rational institutional investor and a utility-driven retail consumer.  We analyze network dynamics across two behavioral regimes. In \textit{The Unbounded Accumulation Model}, the consumer purely accumulates tokens, creating an exclusive buy-side pressure that interacts with institutional portfolio rebalancing to fuel an ever-expanding speculative bubble and generate compounding excess returns for investors. Conversely, in \textit{The Utility-Consumption Model}, the consumer dynamically buys and sells tokens to balance crypto wealth against real-world fiat consumption. Within this framework, we derive an explicit steady-state equilibrium price for ETH, demonstrating how token valuation anchors to a stable fundamental baseline that scales directly with network adoption while completely dissolving the institutional yield premium. Our numerical simulations show that while exogenous traditional finance (TradFi) shocks propagate through portfolio rebalancing to drive high token price volatility, network inflation remains highly stable. Furthermore, we prove that network security is insulated from institutional monopoly by counter-cyclical consumer behavior. Our findings reveal that institutional excess wealth creation in PoS ecosystems is not native to the staking protocol itself, but is strictly driven by the leveraged extraction of the retail consumer's continuous demand for transactional utility.
\end{abstract}

\section{Introduction}

The transition of major decentralized networks—most notably Ethereum—from energy-intensive Proof-of-Work to capital-intensive Proof-of-Stake (PoS) consensus mechanisms has fundamentally intertwined blockchain security with macroeconomic asset pricing theory. In a PoS framework, the cryptographic integrity and Byzantine fault tolerance of the ledger are no longer secured by physical hardware and electricity consumption, but rather by the aggregate financial value of the native token locked directly within the consensus layer. This structural shift gives rise to an endogenous staking yield ($y = c/\sqrt{S}$) that acts as a novel asset class within quantitative finance. While market practitioners frequently evaluate this yield through traditional fixed-income analogies, such static views fail to capture the unique open-economy feedback loops inherent to decentralized protocols. In a dynamic setting, algorithmic token issuance rules, decentralized application utility, and capital rebalancing across traditional and cryptographic frontiers interact continuously to dictate asset pricing, token returns, and systemic security.

To formalize these interactions, this paper presents an open-economy macroeconomic equilibrium model for PoS networks built upon a  dual-agent asset clearing framework. The ecosystem is populated by two "representative" players whose strategic actions determine the state variables: a utility-driven retail Consumer and a Kelly-optimizing institutional Investor. The investor represents an institutional allocator seeking to stake native tokens to hedge a highly volatile traditional financial (TradFi) market. Operating under the assumption of market efficiency, the investor maintains a martingale price expectation where tomorrow's expected price equals today's spot price ($E[p^{t+1}]=p^t$). They maximize the long-term geometric compounding rate of their wealth based on the mean and variance statistics ($\mu_r, \sigma_r^2$) of the traditional market and the protocol yield $y$. Throughout our analysis, we assume that the risk-adjusted TradFi return satisfies $\Delta = \mu_r - \sigma_r^2 < 0$, establishing a powerful variance-drag incentive for institutional capital to hedge into the non-volatile cash flows of the staking protocol. Conversely, the consumer represents regular network users who derive explicit utility from transacting on the ledger. Transactions incorporate the network's contemporary fee-burn architecture (e.g., EIP-1559), meaning that burning tokens for gas permanently destroys both the principal asset and its future compounding cryptographic yield.

The primary contribution of this paper lies in evaluating how different structural regimes of consumer behavior—specifically HODLing versus spending—impact token macroeconomics across two distinct models.

In the first framework, \textit{The Unbounded Accumulation Model}, the consumer class continuously injects a periodic inflow of exogenous fiat capital $I_c$ to buy tokens but never sells their underlying holdings. This foundational HODLing behavior acts as an unyielding source of buying pressure that drives robust structural growth in the token price $p^t$. Because the consumer acts as the exclusive buyer and the investor serves as the sole liquidity provider, the institutional investor's physical token balance ($S_i$) steadily decreases over time as it is systematically transferred to consumer balances ($S_c$). Our numerical results show that this dynamic boosts the return of the native token, generating an ever-expanding speculative bubble where the institutional investor runs a highly profitable, leveraged arbitrage strategy. By continuously absorbing price-insensitive retail demand, the investor converts this utility into a compounding geometric excess return that significantly outperforms the traditional market baseline.

In the second framework, \textit{The Utility-Consumption Model}, the consumer's decision rule expands via a two-level log-linear Cobb-Douglas utility function to encompass both buying and selling dynamics. The consumer dynamically liquidates a fraction ($\nu$) of their wealth pool for real-world fiat consumption while retaining the remainder ($\xi = 1-\nu$) in crypto savings. This two-way asset clearing completely alters the macroeconomic equilibrium. Instead of fueling an ever-expanding bubble, the consumer's active profit-taking and buying behavior anchors the token price to a stable, deterministic fundamental baseline value that scales linearly with network adoption ($N$). Crucially, under this consumption regime, the institutional investor's structural alpha completely vanishes. The cross-sectional distribution of the investor's cumulative excess return collapses symmetrically around zero, demonstrating that their speculative risk becomes entirely uncompensated.

Across both models, our simulations answer critical structural questions regarding PoS economics. First, we show that net token inflation remains highly stable, averaging approximately 1\% annualized over a 10-year horizon ($T=120$). Second, the model accounts for the high historical volatility of crypto assets by demonstrating how exogenous TradFi return shocks ($R^t$) propagate into spot token prices through the investor's continuous portfolio rebalancing. Finally, we show that network security is natively insulated from total collapse. Because the institutional investor is forced to sell native assets during price expansions to maintain their target portfolio weight $w^t$, they never gain monopolistic control over the token supply. Instead, the system is governed by a native "consumer feedback control": asset depreciation amplifies the purchasing power of steady retail fiat inflows, triggering counter-cyclical buying pressure that stabilizes the physical staked supply ($S$). Ultimately, these findings reveal a core macroeconomic reality of Proof-of-Stake networks: institutional excess wealth creation is not an intrinsic property of the staking protocol itself, but is strictly driven by the leveraged extraction of the retail consumer's continuous demand for transactional utility.

\subsection{Literature Review}

This paper directly extends the macroeconomic framework developed in \cite{Perepelitsa2026_I} by relaxing the assumptions of frictionless institutional capital flows and a pure HODLing retail sector, thereby explicitly incorporating endogenous price friction and dynamic consumer liquidations. 

Our framework departs fundamentally from the seminal closed-ecosystem paradigm established by Chitra \cite{Chitra2020}, which isolates capital competition strictly within token-denominated DeFi lending markets and concludes that network security requires artificial algorithmic interest-rate parity to prevent catastrophic capital flight. We extend and reshape this framework in three critical macroeconomic dimensions: first, by constructing an open-economy fiat interface where consensus layer security is determined by direct competition against exogenous traditional finance (TradFi) returns and retail demand for real-world liquidity; second, by formalizing contemporary fee-burn architectures (such as EIP-1559), which treat transaction execution as the permanent destruction of both the principal asset and its future compounding staking yield, rather than temporary capital preservation; and third, by proving mathematically that a stabilizing equilibrium exists natively via endogenous price dynamics and consumer wealth effects, ensuring network security without the need for protocol-enforced interest pegs.

By framing the protocol as an open system, this work contributes to a broader body of literature examining blockchain macroeconomics and token valuation. Early foundations established the structural viability of PoS consensus without hardware externalities \cite{saleh2021blockchain}, while subsequent tokenomics research formalized optimal block rewards \cite{fanti2019economics, john2021equilibrium}, utility-token valuation under velocity constraints \cite{cong2021tokenomics}, and mechanism design for congested transaction fee markets \cite{roughgarden2021transaction}. Our analysis specifically reopens the classical debate surrounding validator wealth concentration; whereas standard closed-system models find that staking rewards do not inherently lead to an unchecked, monopolistic concentration of shares among the wealthiest nodes \cite{rosu2021evolution, leporati2023studying}, our model demonstrates a counter-intuitive reversal when cross-boundary capital rebalancing is allowed. Structurally, we build our models on the recent macro-liquidity and optimal token issuance frameworks developed by Jermann \cite{jermann2025optimal, jermann2025macro}, extending them into an asset-clearing environment with capital frictions.

\section{Preliminaries}

\begin{table}[htbp]
    \centering
    \renewcommand{\arraystretch}{1.6} 
    \begin{tabular}{l p{12cm}}
        \toprule
        \textbf{Symbol} & \textbf{Description} \\
        \midrule
        $M$ & Total fixed physical supply of the native token (e.g., ETH).
\\
        $S$ & Total physical tokens staked in the network.
\\
        $p$ & Fiat-denominated unit price of the native token (e.g., USD price of ETH).
\\
        $y$ & Native staking yield, defined as $y = \frac{c}{\sqrt{S}}$, where $c$ is the network issuance parameter.
\\
        $W_i$ & Total fiat-denominated wealth of the Investor class (combining both TradFi cash and crypto allocations).
\\
        $W_c$ & Fiat-denominated budget of the Consumer class.
\\
        $w(S)$ & The Kelly-optimized portfolio fraction allocated by investors to staking, defined as $w(S) = \frac{\frac{c}{\sqrt{S}} - \Delta}{\sigma_r^2}$, where $\Delta = \mu_r - \sigma_r^2$ is the fiat opportunity cost.
\\
        $\gamma$ & The Consumer's fiat-denominated utility preference for transactional liquidity.
\\
        $U$ & The Consumer's utility function, defined as $U = E[p] S_c(1+y) + \gamma \ln(L_c)$, representing the balance between the expected future fiat value of staked tokens ($S_c$) and the immediate transactional utility of liquid gas ($L_c$).
\\
        $S_i$ & Physical tokens demanded for staking by Investors.
\\
        $S_c$ & Physical tokens demanded for staking by Consumers.
\\
        $L_c$ & Physical tokens held liquid by Consumers for transactional utility.
\\
        \bottomrule
    \end{tabular}
    \caption{State variables and parameters for the open-economy model.}
    \label{tab:notation}
\end{table}

In this section, we establish the main structural components, agent definitions, and parametric assumptions that govern our open-economy macroeconomic framework. The model is populated by two representative actors whose interactions determine the evolution of the token and fiat asset markets: the Consumer and the Investor.

We define the following state variables and parameters for the open-economy model in Table \ref{tab:notation}.

\textbf{The Representative Consumer:} 
The Consumer represents the regular users of the ETH network who derive explicit utility from both staking native assets and executing transactions on the network. A defining operational characteristic of this class is that they provide a periodic, continuous inflow of exogenous fiat capital $I_c$ into the cryptographic ecosystem. Critically, this deployment of outside fiat capital occurs without the immediate selling or liquidation of their underlying token holdings, corresponding directly to structural HODLing behavior. Depending on the institutional constraints and behavioral assumptions imposed on the economy, this capital injection maps to two alternative modeling environments evaluated throughout this paper. In the first setting, termed The Unbounded Accumulation Model, the continuous fiat influx represents pure capital accumulation where consumer wealth is permanently committed to acquiring and locking native tokens within the network protocol. In the second setting, defined as the Utility-Consumption Model, the consumer's decision rule expands to encompass both buying and selling dynamics, allowing them to dynamically balance their crypto allocations against real-world fiat consumption needs to anchor the network to a stable equilibrium state.

\textbf{The Representative Investor:}
The Investor represents a sophisticated institutional allocator who considers investing in ETH staking primarily as a vehicle for hedging a highly volatile traditional financial (TradFi) market. The investor operates under the strict belief of market efficiency, assuming that short-term speculative price momentum is fundamentally unforecastable. Consequently, when formulating their optimal asset allocation and computing target portfolio fractions, the investor assumes that the token market satisfies a martingale price expectation where tomorrow's expected price equals today's spot price. Operating within this efficient market paradigm, the investor utilizes Kelly portfolio optimization to maximize the expected long-term geometric compounding rate of their total wealth. This intertemporal wealth maximization process relies strictly on the observed mean and variance statistics ($\mu_r, \sigma_r^2$) of the TradFi market and the algorithmically determined network staking yield.

\textbf{Parametric Condition on Traditional Returns:}
To characterize an economic environment where crypto-asset staking serves an essential portfolio-hedging function, we restrict our analysis to macroeconomic regimes that satisfy the following structural condition:
\begin{equation*}
    \Delta = \mu_r - \sigma_r^2 < 0
\end{equation*}
Economically, this parameter condition implies that the inherent variance drag of the outside traditional finance market dominates its arithmetic mean return. This net negative baseline establishes a continuous macroeconomic incentive for risk-averse institutional capital to reallocate a portion of its portfolio away from traditional assets and hedge into the stable, non-volatile token rewards algorithmically guaranteed by the staking consensus layer.

\section{The Unbounded Accumulation Model}

\subsubsection{The Opportunity Cost of Consumed Liquidity}
\label{sec:opportunity_cost}

In traditional macroeconomic models of fiat currency, such as Money-in-the-Utility (MIU) frameworks, liquid cash is held rather than consumed.
Consequently, the opportunity cost of holding liquidity is strictly the foregone yield, $y$.
However, the Ethereum network's fee-burn mechanism (EIP-1559) requires a structural departure from this assumption.
Tokens utilized for transaction fees are permanently removed from the supply.
Therefore, the true opportunity cost of liquidity must account for the destruction of both the principal token and its future compounding yield.
Note, that this is different from the assumptions of \cite{Perepelitsa2026_I} and \cite{jermann2025macro, jermann2025optimal}.

We can formulate the consumer's decision as a balance between the subjective marginal benefit of burning a token and the objective market cost of foregone staking.
Let the consumer's utility derived from consumed liquidity (gas), $L_c$, be scaled by a preference parameter $\gamma$, such that the utility of transaction execution is $\gamma \log(L_c)$.
The marginal utility of an additional liquid token is $\frac{\gamma}{L_c}$.
Evaluated at the current fiat token price $p$, the subjective marginal benefit per physical token is:

$$ \text{Marginal Benefit} = \frac{\gamma}{p L_c} $$

This term represents the fiat-adjusted value the consumer places on burning one additional physical token in the current period.
Conversely, to acquire that additional liquid token, the consumer must sacrifice exactly one staked token, $S$.
In the Proof-of-Stake market, a single staked token automatically compounds to $1+y$ physical tokens in the subsequent period.
Thus, the true physical cost of burning a token today is the permanent loss of both the principal and its future interest:

$$ \text{Marginal Cost} = \frac{E[p](1 + y)}{p}, $$
where $E[p]$ is the expected price of a token at the next period.
A rational consumer optimizes their allocation by equating the subjective marginal benefit of consumption with the objective opportunity cost of staking.
In equilibrium, the consumer ceases to shift tokens from staking to liquidity exactly when:

$$ \frac{\gamma}{p L_c} = \frac{E[p](1 + y)}{p}.$$

Solving for the optimal liquid demand yields:

\begin{equation}
\label{eq:L}
L_c = \frac{\gamma}{E[p](1+y)}.
\end{equation}
Futhermore, to simplify the model, we will assume that consumers best estimate today is that the expected tomorrow's price equals today's price: $E[p]=p.$

Now we turn to the construction of a dynamic market model where the variables evolve from one time period $t$ to the next $t+1,$ as the total supply of ETH ($M^t$) changes due to ETH issuance and burn and the TradFi returns fluctuate randomly, triggering changes in other variables.

\subsubsection{Network supply evolution $M^{t+1}$}
At period $t,$ the total supply of tokens $M^t$ consisists of the staked tokens $S^t$ and tokens that will burned over one time step $L^t_c$:
\[
M^t = S^t + L^t_c.
\]
Then, at the next step
\[
M^{t+1} = S^t(1+y^t).
\]
Similarly, by designating the new amounts to stake and to burn, the new supply also equals
\[
M^{t+1} = S^{t+1} + L^{t+1}_c,
\]
where $L^{t+1}$ is the new demand for gas by the consumers.
From the last two equations, the new staking level is related to the last period by equation 
\begin{equation}
\label{eq:S_t+1}
S^{t+1} = S^t(1+y^t) - L^{t+1}_c.
\end{equation}

\subsubsection{Next period price $p^{t+1}$}
The new price is set to match the tokens that the Investor frees after rebalancing with the amount demanded by the Consumer. For that, we first compute the wealth accumulated by the Investor and the Consumer over one period:
\begin{itemize}
    \item Investor Outside Cash (random TradFi return): 
    \[
    K^{t+1}_i = W_i^t (1 - w^t)(1 + R^t),
    \]
    where $R^t$ is a random variable with mean $\mu_r$ and variance $\sigma_r^2.$
    \item Investor Total Wealth: 
    \begin{equation}
\label{eq:W_i new}
    W_i^{t+1} = K^{t+1}_i + p^{t+1} \frac{w^t W_i^t}{p^t} (1 + y^t){}={} K^{t+1}_i + p^{t+1}S^t_i(1+y^t).
\end{equation}

    \item Consumer Total Wealth ( including exogenous fiat income $I_c^{t+1}$): 
    \begin{equation*}
    \label{eq:W_C new}
    W_c^{t+1} = p^{t+1} \left( \frac{W_c^t}{p^t} - L_c^t \right) (1 + y^t) + I^{t+1}_c.
\end{equation*}
\end{itemize}
The Investor designates $w^{t+1}W_i^{t+1}$ in fiat to stake at the next period. The change in the number of tokens held by 
the Investor equals:
\[
S^{t}_i(1+y^t) - \frac{w^{t+1}W_i^{t+1}}{p^{t+1}}{}={}
(1-w^{t+1})S^t_i(1+y^t){}-{}\frac{w^{t+1}K_i^{t+1}}{p^{t+1}}.
\]
This amount is bought by consumers with the outside income $I_c^{t+1}:$
\[
(1-w^{t+1})S^t_i(1+y^t){}-{}\frac{w^{t+1}K_i^{t+1}}{p^{t+1}}
{}={}\frac{I^{t+1}_c}{p^{t+1}}.
\]
From this equation, we find that
\begin{equation*}
    (1-w^{t+1})S^t_i(1+y^t){}={}\frac{w^{t+1}K^{t+1}_i + I^{t+1}_c}{p^{t+1}},
\end{equation*}
or
\begin{equation*}
\label{eq:new price}
    p^{t+1} = p^t\left[ \frac{w^{t+1} K^{t+1}_i + I_c^{t+1}}{(1 - w^{t+1})w^t W_i^t (1 + y^t)}\right],
\end{equation*}
or
\begin{equation*}
\label{eq:new price 2}
    p^{t+1} = \frac{w^{t+1} K^{t+1}_i + I_c^{t+1}}{(1 - w^{t+1})S_i^t (1 + y^t)}.
\end{equation*}
In deriving the above formula, we assumed that $I^{t+1}_c\geq 0.$
From the last two equations, we also find the balance of tokens owned by investors over one period: 
\begin{equation}
\label{eq:ETH investor}
S_i^{t+1} = S_i^t(1+y^t)- I_c^{t+1}/p^{t+1}.
\end{equation}

\subsubsection{Equation for $S^{t+1}.$}
Substituting the dynamic price $p^{t+1}$ and liquid demand $L^{t+1}_c$ into the physical balance equation \eqref{eq:S_t+1} yields equation for $S^{t+1}$:
\begin{equation*}
\label{eq:Master}
    S^{t}(1+y^t) - S^{t+1} = \frac{\gamma \sqrt{S^{t+1}} \left[ (1 - w^{t+1}) S_i^t (1 + y^t) \right]}{(c+\sqrt{S^{t+1}}) \left( w^{t+1} K^{t+1}_i + I_c^{t+1} \right)}
\end{equation*}
It can be shown that this equation has only one positive root in the interval $(0,\sqrt{M^{t+1}}).$
This value corresponds to the amount of staked ETH at the next level.
Once $S^{t+1}$ is found, $w^{t+1}$, $p^{t+1}$, and the wealth state variables are sequentially updated.

\subsection{Price dynamics}
Using \eqref{eq:W_i new} we can write the gross return as
\begin{equation}
    \label{eq:return basic}
    \frac{p^{t+1}}{p^t} {}={}
    \frac{w^{t+1}}{1-w^{t+1}}\frac{1-w^t}{w^t}\frac{1+R^t}{1+y^t} {}+{}\frac{I_c^{t+1}/W_i^t}{(1-w^{t+1})w^t(1+y^t)}.
\end{equation}
Thus, the price dynamics is driven by the TradFi market
and the in-flow of fiat from the Consumer. 


Likewise, we obtain a formula for the return to the Investor:
\begin{equation}
\label{I:return to investor}
    \frac{W_i^{t+1}}{W_i^t} = (1 + R^t) + \frac{I_c}{1 - w} \frac{1}{W_i^t}.
\end{equation}
This recurrence relation can be solved explicitly:
\[
W_i^t = W_i^0 \prod_{k=0}^{t-1} (1 + R^k) + \frac{I_c}{1 - w} \sum_{j=0}^{t-1} \prod_{k=j+1}^{t-1} (1 + R^k).
\]
The growth of $W^t_i$ can be estimated by assuming that $R^t=R$ is a constant, in which case we get
\[
W_i^t = W_i^0 (1 + R)^t + \frac{I_c}{1 - w} \left[ \frac{(1 + R)^t - 1}{R} \right].
\]
We compare this return with the baseline return from a TradFi account ($W^t_{TradFi}=W_i^0(1+R)^t$). The excess of the cumulative return in the logarithms is
\begin{equation}
\label{est:log excess}
    \log W^t_i - \log W^t_{TradFi} \approx \log \left(\frac{I_c^t}{(1-w)R}\right).
    \end{equation}
 
We can summarize the dynamics as follows. 
When there is no inflow of Consumer fiat $I_c=0,$
the returns of ETH price are simply scaled returns to TradFi. With
periodic additional boost from the Consumer $I_c>0,$ the price of ETH goes up in an ever-expanding bubble, with the volatility  equal to a multiple of the TradFi plus the volatility of the inflow $I_c^t.$

The term $\frac{I_c}{1-w}$ in the return of the investor portfolio in \eqref{I:return to investor} represents the leveraged extraction of ETH demand, captured by the liquidity provider (Investor). Even though the improvement over TradFi in \eqref{est:log excess} is only logarithmic in time, in the short term, it can make a significant contribution to the growth of investor wealth. 

In this dynamic bubble driven by the HODLing Consumer and the Kelly's Investor, the security of the network is not at threat, as the Investor does not gain control over the supply of ETH. According to equation \eqref{eq:ETH investor}, for all realistic values of the yield $y^t,$ the amount of ETH owned by the Investor is decreasing.

In the next section, we illustrate these type dynamics by running a numerical simulation of the ETH economy.
 
\subsection{Numerical Tests}
\begin{figure}[htbp]
    \centering
 \includegraphics[width=0.8\textwidth]{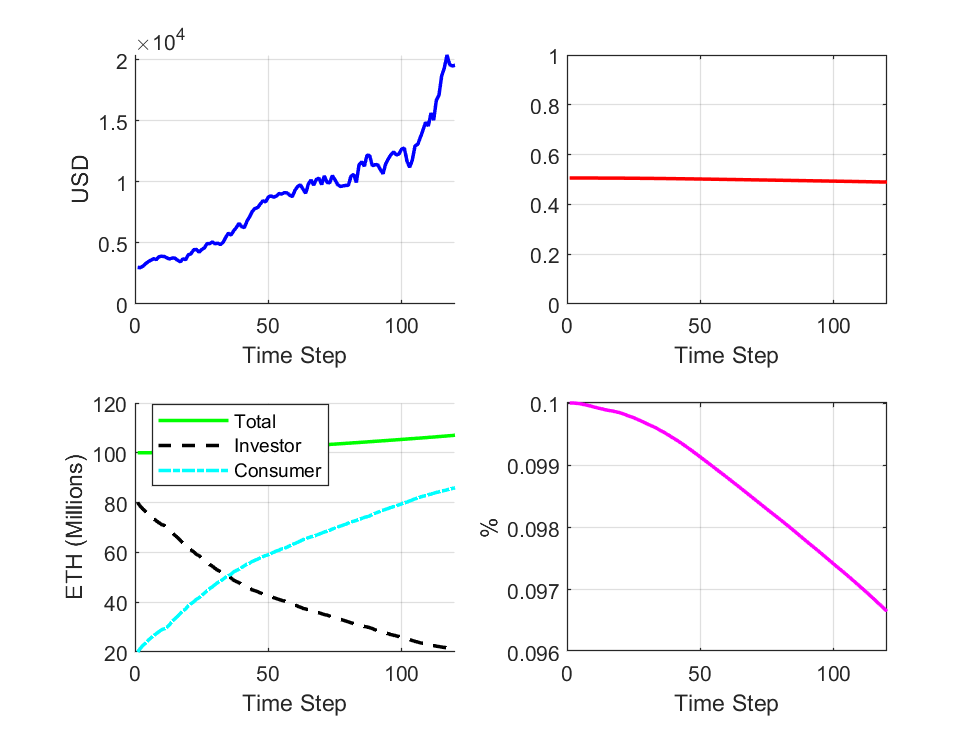} 
    \caption{A single realization of the market dynamics over 120 months in the model with unbounded accumulation. The ETH price $p^t$ (top left), the Investor portfolio fraction $w^t$ (top right), amounts of ETH owned by the Investor (dash) and Consumer (blue), (bottom left), ETH yield $y^t$ (bottom right)}
    \label{fig:model1 dash}
\end{figure}

\begin{figure}[htbp]
    \centering
\includegraphics[width=0.6\textwidth]{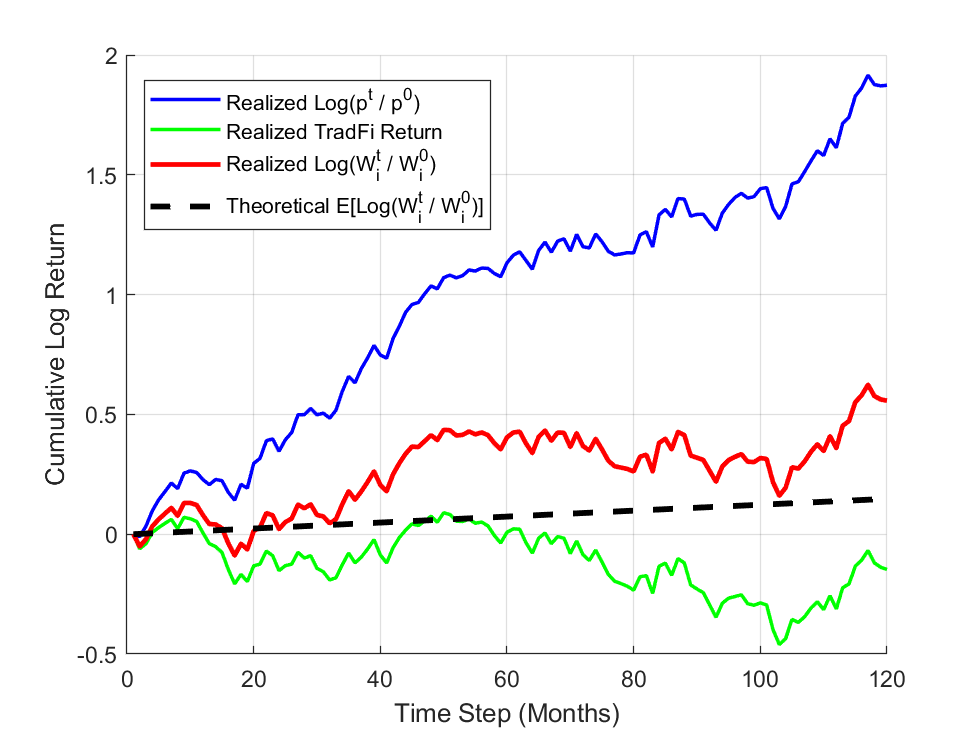} 
    \caption{A single realization of returns over 120 months.}
    \label{fig:model1 returns}
\end{figure}

\begin{figure}[htbp]
    \centering
\includegraphics[width=\textwidth]{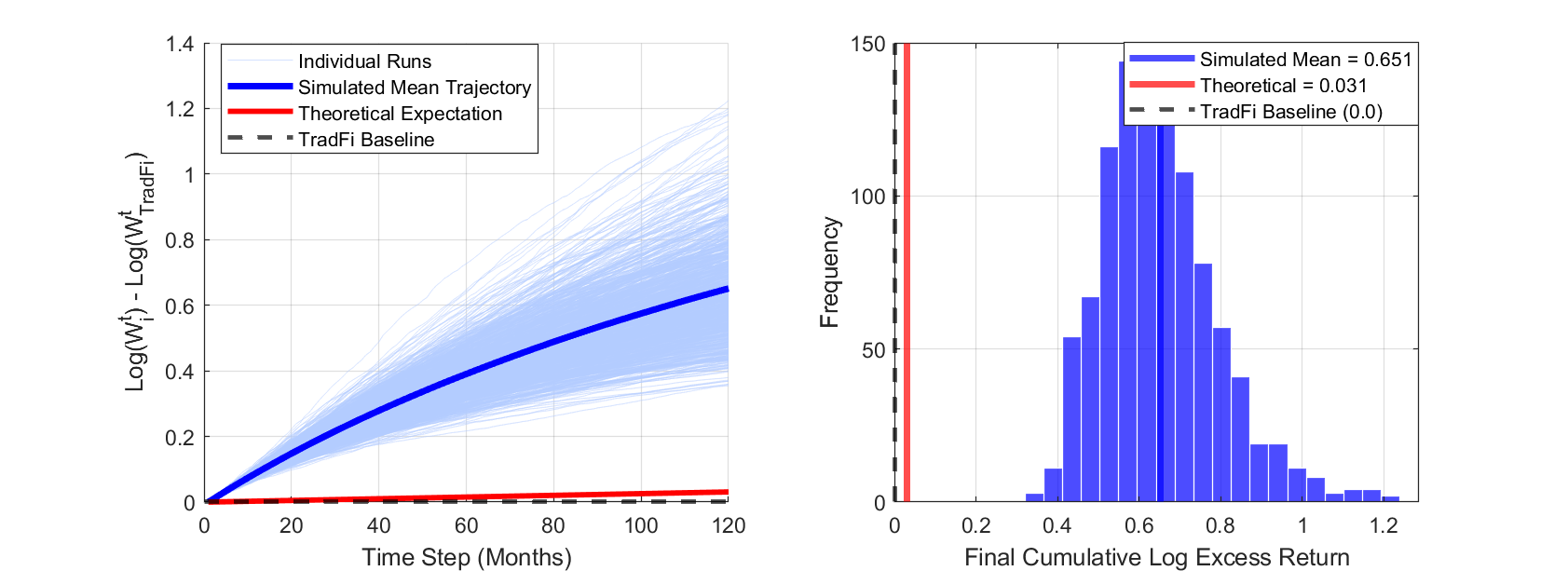} 
    \caption{Statistics of the Investor excess returns over 1000 independent realizations of the market dynamics over 120 months. Plot of 1000 trajectories and the mean (left), distribution of the Investor log excess returns at $T=120$ (right). }
    \label{fig:model1 stats}
\end{figure}

Figures \ref{fig:model1 dash}, \ref{fig:model1 returns} show one realization of the market dynamics, and Figure \ref{fig:model1 stats} shows the statistical information collected over 1000 independent simulations, using the same initial data.

For the numerical simulations, we initialize the model with a time horizon of $T=120$ periods, representing a 10-year monthly trajectory. The traditional finance (TradFi) market is calibrated to a realistic monthly scale, utilizing a mean return of $\mu_r = 0.002$ (0.2\% per month) and a volatility of $\sigma_r = 0.045$ (approximately 15.5\% annualized). The network's issuance parameter is set to $c = 10$, and the consumer's utility preference for gas is fixed at $\gamma = 3 \times 10^8$ USD. The exogenous retail fiat injection ($I_c^t$) is modeled as a white noise process with a long-run mean of $\mu_{I} = 4 \times 10^9$ USD per month and a standard deviation of $\sigma_{I} \approx 0.759 \times 10^9$ USD, floored at $5 \times 10^8$ USD to prevent total consumer depletion. At $t=1$, the system's initial conditions are anchored with a token price of $p^1 = 3000$ USD, a total staked physical supply of $S^1 = 100 \times 10^6$ tokens, and an institutional allocation of $S_i^1 = 80 \times 10^6$ tokens (Investor controls 80\% of ETH).

The numerical simulation reveals distinct macroeconomic interactions between token valuation, network issuance, and capital reallocation. The token price $p^t$ exhibits robust structural growth over the 10-year horizon ($T=120$), driven by the continuous, price-insensitive inflow of retail fiat capital $I_c^{t+1}$. Short-term price volatility is driven by exogenous traditional finance (TradFi) shocks $R^t$, which propagate through the system as the Kelly-optimizing investor dynamically adjusts their portfolio wealth fraction $w^t$ to preserve target risk allocations. Regarding network issuance, the total staked token supply behaves as a highly stable variable, maintaining a smooth annualized net token inflation rate of approximately 1\% over the 10 years.

Ultimately, the cross-sectional Monte Carlo results over 1,000 simulated market realizations show that a simple hedging strategy of the institutional investor in reality plays out as a highly profitable, leveraged arbitrage strategy against the retail class, when the consumer holds on to its tokens.

By continuously absorbing constant demand for utility, the investor captures these fiat injections into TradFi reserves. This structural setup converts retail utility demand into a massive, compounding geometric excess return (achieving a simulated mean excess return of 0.656), vastly outperforming both the TradFi baseline and the investor's own passive, yield-only theoretical expectation. The theoretical returns 
are defined using the geometric expectations:
\[
W_{theor}^{t+1}/W_{theor}^t{}={}e^{g^t},
g^t{}\approx{}w^t y^t + (1 - w^t)\mu_r - \frac{1}{2}(1 - w^t)^2\sigma_r^2.
\]







\section{The Utility-Consumption Model}

\subsection{Consumer 2-level utility}
We consider a logarithmic Cobb-Douglas utility function to determine the consumer preferences for selling ETH $C,$ the amount of staked ETH $S_c,$ and the gas ETH $L_c$ to be
\begin{equation}
    U_1(C,V){}={}
    \nu\log C{}+  (1-\nu)\log V,\quad \nu\in[0,1],
\end{equation}
with the budget constraint 
$$C+V{}={}W_c.$$
The second level utility function is used to determined the preferences for ETH gas $L_c$ and staked ETH $S_c,$ given the crypto fraction of the wealth and the inflow of consumer fiat for buying ETH $I_c:$
\begin{equation}
    U_2(S_c,L_c) = E[p]S_c(1+y) + \gamma\log L_c,
\end{equation}
with the budget
\[
pS_c + pL_c = V.
\]
Here, $E[p]$ is the next period ETH price.
Optimizing the consumption we get:
\[
C= \nu W_c,\quad
V = \xi W_c,\quad \xi=1-\nu.
\]

Optimizing the crypto balance, we obtain:

\[
L_c=\frac{\gamma}{E[p](1+y)},
\]
\[
S_c = V/p- L_c {}={}\frac{\xi W_c}{p}-L_c.
\]
With this designation of the capital, the wealth of the consumer will change to 
\[
W_c^{new}{}={}E[p]\left(\xi W_c/p -L_c\right)(1+y) + I_c,
\]
where $I_c$ is an exogenous inflow of consumer fiat.

\subsubsection{Network supply evolution $M^{t+1}$}
As in the first model, we have the following equations.
The total supply of ETH at period $t:$
\[
M^t = S^t + L^t_c.
\]
The number of tokens changes due to the network inflation and burn:
\[
M^{t+1} = S^t(1+y^t).
\]
The new amount of tokens staked is determined from equation
\[
M^{t+1} = S^{t+1} + L^{t+1}_c,
\]
where $L^{t+1}$ is the new demand for gas by the consumers.
And finally,
\begin{equation}
\label{eq:S_t+1 II}
S^{t+1} = S^t(1+y^t) - L^{t+1}_c.
\end{equation}


\subsubsection{Next period price $p^{t+1}$}

The wealth accumulated by the Investor and the Consumer over one period:
\begin{itemize}
    \item Investor Outside Funds (random TradFi return): 
    \[
    K^{t+1}_i = W_i^t (1 - w^t)(1 + R^t),
    \]
    where $R^t$ is a random variable with mean $\mu_r$ and variance $\sigma_r^2.$
    \item Investor Total Wealth: 
    \begin{equation}
\label{eq:W_i new 2}
    W_i^{t+1} = K^{t+1}_i + p^{t+1} \frac{w^t W_i^t}{p^t} (1 + y^t){}={} K^{t+1}_i + p^{t+1}S^t_i(1+y^t).
\end{equation}

    \item Consumer Total Wealth (including exogenous fiat income $I_c^{t+1}$): 
    \begin{equation*}
    \label{eq:W_C new 2}
    W_c^{t+1} = p^{t+1} \left( \frac{\xi W_c^t}{p^t} - L_c^t \right) (1 + y^t) + I^{t+1}_c.
\end{equation*}
\end{itemize}

The new price $p^{t+1}$ clears the demand for tokens (right) with the total supply of tokens (right):
\begin{equation}
\label{eq:token balance}
S^t(1+y^t){}={}\frac{w^{t+1}W^{t+1}_i + \xi W_c^{t+1}}{p^{t+1}}.
\end{equation}
Since,
\[
S^t(1+y^t){}={}\frac{w^tW^t_i(1+y^t)}{p^t} +
\left(\frac{\xi W^t_c}{p^t} - L^t_c\right)(1+y^t),
\]
we can express \eqref{eq:token balance} as a balance of sold (left) and bought (right) tokens:
\begin{equation*}
\left(
(1-w^{t+1})w^t W_i^t + \nu\left(\xi W_c^t - p^tL^t_c\right) \right)\frac{1+y^t}{p^t} 
{}={}   
\frac{w^{t+1}K_i^{t+1} +\xi I_c^{t+1}}{p^{t+1}}.
\end{equation*}
This equation leads to the price change equation
\begin{equation*}
\label{eq:new price II}
    p^{t+1} = p^t\left[ 
    \frac{w^{t+1}K_i^{t+1} +\xi I_c^{t+1}}
    {\left(
(1-w^{t+1})w^t W_i^t + \nu\left(\xi W_c^t -p^tL^t_c\right) \right)(1+y^t)}
\right],
\end{equation*}
and using the formula for $L^t_c={}={}\frac{\gamma}{p^t(1+y^t)},$ we obtain:
\begin{equation*}
\label{eq:new price III}
    p^{t+1} = p^t\left[ 
    \frac{w^{t+1}K_i^{t+1} +\xi I_c^{t+1}}
    {\left(
(1-w^{t+1})w^t W_i^t + \nu\left(\xi W_c^t -\frac{\gamma}{1+y^t}\right) \right)(1+y^t)}
\right],
\end{equation*}
or
\begin{equation*}
\label{eq:new price IV}
    p^{t+1} = p^t\left[ 
    \frac{w^{t+1}(1-w^t)W_i^t(1+R^t) +\xi I_c^{t+1}}
    {\left(
(1-w^{t+1})w^t W_i^t + \nu\xi W_c^t \right)(1+y^t) - \nu\gamma}
\right].
\end{equation*}
As in the previous model, using the new price $p^{t+1}$
in equation \eqref{eq:S_t+1 II}, we obtain a non-linear equation for the new staking amount $S^{t+1}.$
The equation determined a unique solution in the range $(0,M^{t+1}).$ Once $S^{t+1}$ is determined, all remaining variables are computed from the above formulas.

\subsection{Special case: a pure consumer market}
\label{sec:pure_consumer_equilibrium}
To understand the intrinsic fundamental value of the network, we analyze the limiting case where the traditional finance sector is entirely absent. In this pure peer-to-peer retail economy, the institutional investor's initial wealth is zero ($W_i^0 = 0$). Consequently, the institutional sector cannot act as a market maker, and the retail consumer class must hold the entirety of the physical staked token supply, such that $S_c = S$.

We begin with the law of motion for the consumer's total fiat-denominated wealth, $W_c^{t}$. In the dynamic model, the consumer's wealth updates according to their retained crypto allocation and their exogenous fiat income $I_c^{t+1}$:
\begin{equation}
\label{eq:W_c eq}
    W_c^{t+1} = p^{t+1} \left( \frac{\xi W_c^t}{p^t} - L_c^t \right) (1+y^t) + I_c^{t+1}
\end{equation}
where $\xi \in (0,1)$ is the consumer's crypto retention fraction, $p$ is the token price, $y = c/\sqrt{S}$ is the staking yield, and $L_c^t$ represents the physical tokens burned to satisfy the fiat-denominated gas utility parameter $\gamma$. The gas demand constraint is defined as:
\begin{equation*}
    L_c^t = \frac{\gamma}{p^t (1+y^t)}
\end{equation*}

The balance of tokens at the new period is:
\[
p^{t+1} \left( \frac{\xi W_c^t}{p^t} - L_c^t \right) (1+y^t) =
\xi W_c^{t+1}.
\]
Using this equation in \eqref{eq:W_c eq} we find the new wealth of the consumer is determined by the cash inflow $I_c^{t+1}:$
\begin{equation*}
    \label{eq:W_c eq 2}
    W_c^{t+1}{}={}\frac{I_c^{t+1}}{\nu}.
\end{equation*}
Using this back in equation \eqref{eq:W_c eq}, we find the price of the token:
\begin{equation*}
    \label{eq:p^t eq}
    \frac{p^{t+1}}{p^t} = \frac{\xi I_c^{t+1}}{\xi I_c^{t}(1+y^t) - \nu \gamma}.
\end{equation*}
The new period staking $S^{t+1}$ is determined from the non-linear equation
\begin{equation*}
    \label{eq:S^t eq}
    S^t(1+y^t){}={}  S^{t+1} + \frac{\gamma}{p^{t+1}(1+y^{t+1})},\quad y^{t+1}{}={}\frac{c}{\sqrt{S^{t+1}}}.
\end{equation*}
\subsubsection{Steady state}
This consumer dynamics admits a unique steady state if we assume that the fiat inflow is constant: $I_c^t=I_c.$ In such a case, the equilibrium values for the price, the consumer wealth, the staking amount, and ETH yield can be obtained from the above equations. The values are as follows.
\[
W_c = \frac{I_c}{\nu},\quad  S^* = \left( \frac{c \xi I_c}{\gamma \nu} \right)^2,\quad p = \frac{\gamma^2 \nu}{c^2 (\xi I_c + \gamma \nu)}, \quad y  = \frac{\gamma \nu}{\xi I_c}.
\]
It can be shown that this state is stable for the discrete-time dynamics described above. 

The steady-state gives a baseline price of ETH, expressing the "usefulness" of the token:
\begin{equation}
\label{eq:price eq}
p = \frac{\gamma^2 \nu}{c^2 (\xi I_c + \gamma \nu)}.
\end{equation}

In the market consisting of $N$ consumers, the value of $\gamma$ is aggregated over the consumers, so that $\gamma=\gamma_0 N,$ where $\gamma_0$ is the individual preference of transact in ETH. The periodic fiat inflow is also aggregated:  $I_c=I_0 N.$
Substituting these into the formula for $p,$ we find that the baseline price grows with increasing adoption of ETH, linearly in $N.$

\subsection{Numerical tests}

\begin{figure}[htbp]
    \centering
 \includegraphics[width=0.8\textwidth]{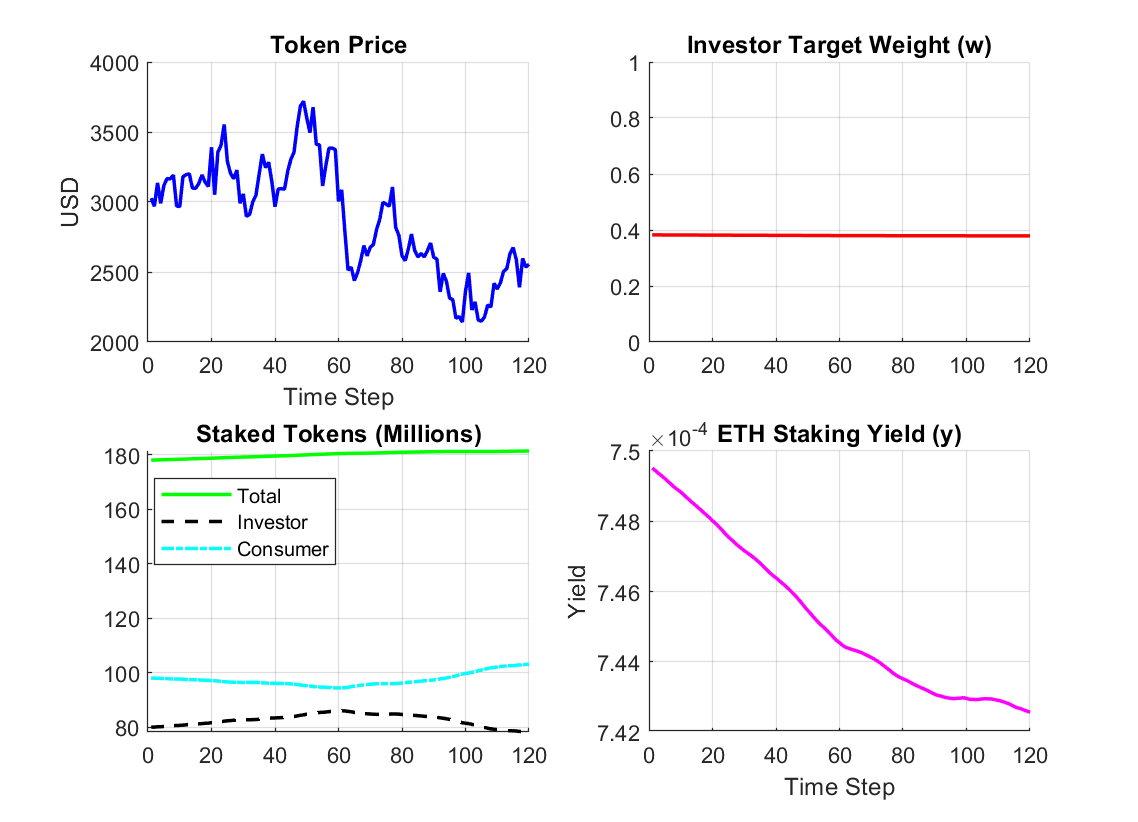} 
    \caption{A single realization of the market dynamics over 120 months in the model with utility consumption. The ETH price $p^t$ (top left), the Investor portfolio fraction $w^t$ (top right), amounts of ETH owned by the Investor (dash) and Consumer (blue), (bottom left), ETH yield $y^t$ (bottom right)}
    \label{fig:model2 dash}
\end{figure}

\begin{figure}[htbp]
    \centering
\includegraphics[width=0.6\textwidth]{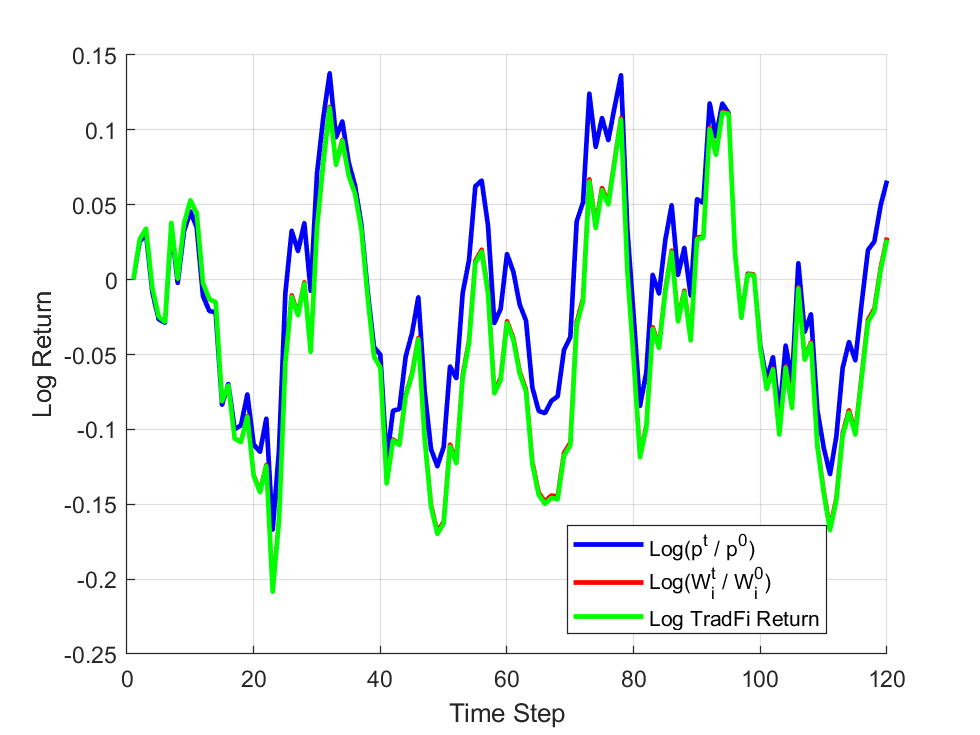} 
    \caption{A single realization of returns over 120 months in the model with utility consumption.}
    \label{fig:model2 returns}
\end{figure}

Figures \ref{fig:model2 dash}, \ref{fig:model2 returns} show one realization of the market dynamics, and Figure \ref{fig:model2 stats} shows the statistical information collected over 1000 independent simulations, using the same initial data.

The model is simulated on a monthly scale over a long-term horizon with a continuous, constant injection of retail capital.

The simulation spans 120 periods, which represents a 10-year monthly timeline. In the traditional finance market, we assume a mean monthly return of 0.2\% and a monthly volatility of 4.5\%, translating to roughly 15.5\% annualized. The traditional finance opportunity cost is calculated as the mean return minus the variance to represent the risk-adjusted baseline. On the network side, the issuance parameter is set to 10. The retail sector injects a constant 3 Billion USD per month, and the gas utility parameter is fixed at $\gamma=300$ million to represent the baseline fiat-adjusted demand for network transactions. Consumer preferences are modeled with a fiat consumption weight of $\nu=0.01$ and a crypto savings weight of $1-\nu=0.99$. This creates a behavior profile where consumers allocate 1\% of their wealth to active fiat consumption and retain 99\% in crypto wealth.

To prevent artificial initialization shocks, the consumer sector and native token price are anchored exactly to their theoretical deterministic steady-state values. The initial consumer wealth is locked at 300 Billion USD. Both the initial token price and the initial consumer staked supply are set to their equilibrium levels, determined by the system's utility parameters and the equilibrium staking yield through equation \eqref{eq:price eq}.

The institutional investors begin with a baseline cryptographic capital allocation of 80 million staked tokens. The total staked tokens in the network combine this investor supply with the consumer supply, which directly determines the initial system staking yield. Using this initial yield, the investor portfolio weight is Kelly-optimized and strictly bounded between 0.001 and 0.999. The initial institutional wealth is then dynamically capitalized to ensure the institutional class begins in perfect portfolio equilibrium based on their target weight. 

Finally, the initial gas liquidity is set according to the optimal liquid demand matching the initial token price and system yield.

\begin{figure}[htbp]
    \centering
\includegraphics[width=\textwidth]{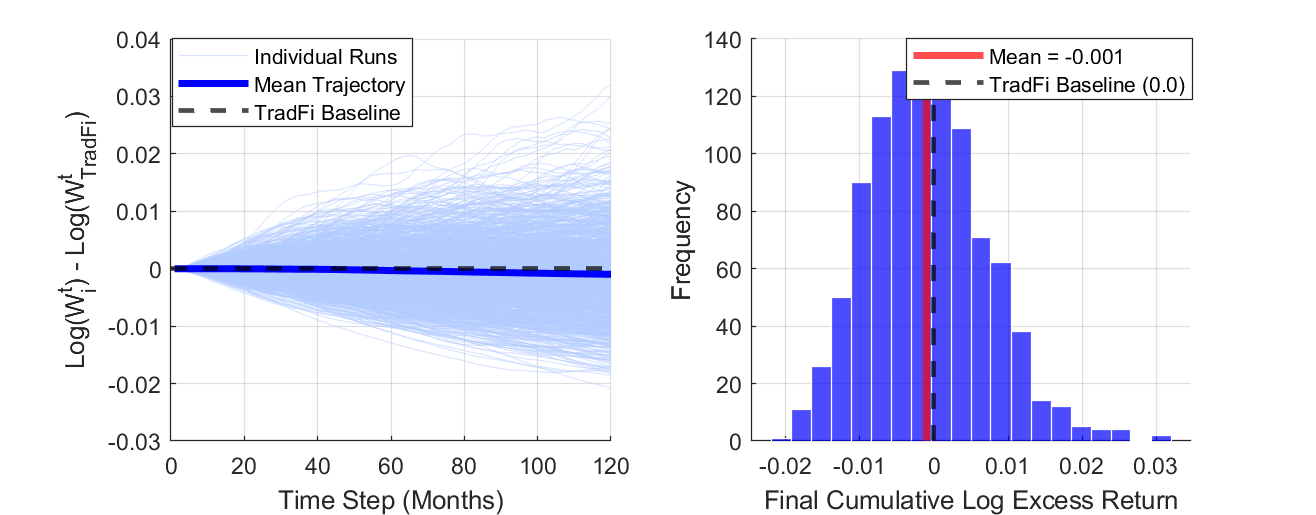} 
    \caption{Statistics of the Investor log excess return over 1000 independent realizations of the market dynamics over 120 months. Plot of 1000 trajectories and the mean (left), distribution of the final cumulative log excess return distribution at $T=120$ (right). }
    \label{fig:model2 investor stats}
\end{figure}

Figure \ref{fig:model2 investor stats} presents the Monte Carlo simulation of the institutional investor's cumulative log excess return ($\ln(W_i^t) - \ln(W_{\text{TradFi}}^t)$) over 1000 simulated market realizations. In stark contrast to the structural outperformance observed in the first model, the mean expected trajectory in this scenario remains near the zero-line. 

The individual market trajectories fan out extensively both above and below the zero-line, showing that while the investor is exposed to severe crypto-specific speculative volatility, this risk is entirely uncompensated. The expected structural premium vanishes completely.

The cross-sectional distribution of the final cumulative excess return at $T=120$ (Figure \ref{fig:model2 investor stats}, right panel) confirms this zero-sum reality. The probability distribution is centered symmetrically around a mean of $-0.001$, which is statistically indistinguishable from the zero TradFi baseline.



\begin{figure}[htbp]
    \centering
\includegraphics[width=\textwidth]{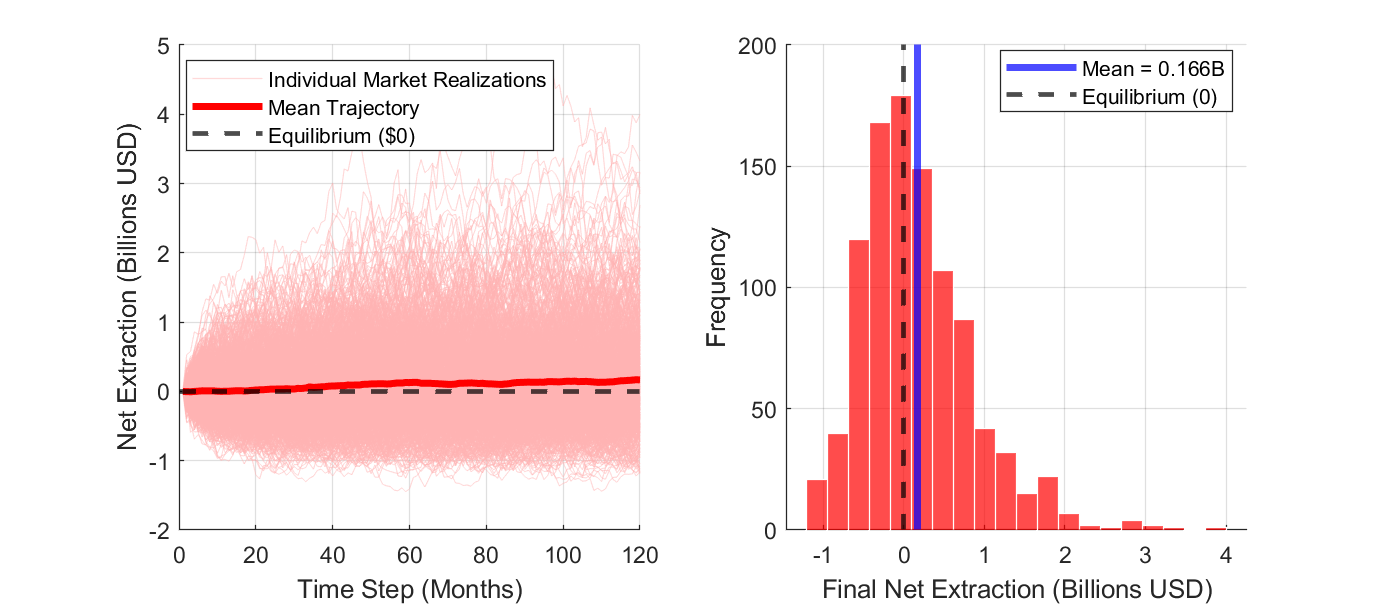} 
    \caption{Statistics of the Consumer net fiat extraction over 1000 independent realizations of the market dynamics over 120 months. Plot of 1000 trajectories and the mean (left), distribution of the Consume net fiat extraction at $T=120$ (right). }
    \label{fig:model2 consumer stats}
\end{figure}

Figure \ref{fig:model2 consumer stats} illustrates the Monte Carlo simulation of the consumer class's net fiat extraction, defined as the difference between their real-world fiat consumption and their exogenous fiat injections ($\nu W_c^t - I_c^t$). As shown in the trajectory panel, while individual market realizations exhibit massive volatility due to stochastic TradFi shocks propagating through the token price, the mean expected trajectory remains close to zero. This empirical result perfectly validates the theoretical steady state ($W_c^* = I_c/\nu$). Economically, the consumer absorbs market chaos: token price appreciation triggers proportional fiat profit-taking (net selling), while price depreciation amplifies the relative purchasing power of their steady fiat income (net buying), constantly forcing the network back toward its fundamental equilibrium. 

The cross-sectional distribution of net extraction at $T=120$ (Figure \ref{fig:model2 consumer stats}, right panel) further confirms this structural stability. The modal density is heavily concentrated exactly at the zero-bound, proving that the system does not structurally drift away from the steady state over long time horizons. The slight positive skew in the mean (e.g., 0.17 Billion USD) is not a deviation from equilibrium, but rather a natural statistical artifact of geometric asset pricing; because fiat-denominated token prices are floored at zero but theoretically uncapped on the upside, rare but extreme stochastic bull markets generate long right-tail asymmetric wealth effects. 



\section*{Acknowledgments}

During the preparation of this work, the author used Gemini 3.1 Pro (Google) to assist with LaTeX formatting, Matlab simulation, structuring the literature review, and refining the academic prose of specific sections. After using this tool, the author reviewed and edited the content as needed and takes full responsibility for the mathematical proofs, models, and final content of the publication.

\clearpage

\bibliographystyle{plain}
\bibliography{refs}

\end{document}